\documentclass[11pt,a4paper]{article}
\usepackage[utf8]{inputenc}
\usepackage{graphicx,verbatim,array,multicol,courier}
\usepackage{amsmath,amssymb,amsthm,mathrsfs, mathtools}
\usepackage{color, graphics, graphicx, siunitx}
\usepackage{amsfonts, dsfont, bm}
\usepackage{natbib} 
\usepackage{arydshln, topcapt}
\usepackage{geometry}
\usepackage{enumitem,float}
\usepackage{lscape}
\usepackage[pdftex,bookmarks,colorlinks]{hyperref}
\usepackage[dvipsnames]{xcolor}
\usepackage{tabularx}
\usepackage{siunitx}
\usepackage[labelformat=simple]{subfig}
\usepackage[linesnumbered,ruled,vlined]{algorithm2e}
\usepackage{booktabs}

\newlist{inparaenum}{enumerate}{2}
\setlist[inparaenum,1]{label=(\alph*)}
\setlist[inparaenum,2]{label=(\roman{inparaenumi}\emph{\alph*})}

\makeatletter
\def\adl@drawiv#1#2#3{%
        \hskip.5\tabcolsep
        \xleaders#3{#2.5\@tempdimb #1{1}#2.5\@tempdimb}%
                #2\z@ plus1fil minus1fil\relax
        \hskip.5\tabcolsep}
\newcommand{\cdashlinelr}[1]{%
  \noalign{\vskip\aboverulesep
           \global\let\@dashdrawstore\adl@draw
           \global\let\adl@draw\adl@drawiv}
  \cdashline{#1}
  \noalign{\global\let\adl@draw\@dashdrawstore
           \vskip\belowrulesep}}
\makeatother

\numberwithin{equation}{section}
\theoremstyle{definition}
\newtheorem{defi}{Definition}[section]
\newtheorem{cond}[defi]{Condition}
\theoremstyle{plain}
\newtheorem{theo}[defi]{Theorem}

\newtheorem{cor}[defi]{Corollary}

\theoremstyle{remark}

\theoremstyle{example}

\newcommand{\To}{\mbox{\upshape\bfseries to}}

\newcommand{\diff}{\mathrm{d}}

\newcommand{\ind}{\mathbbm{1}}

\def\ind{ {{\rm 1}\hskip-2.2pt{\rm l}}}
\definecolor{navy}{rgb}{0,0,0.502}
\definecolor{brown}{rgb}{0.59, 0.29, 0.0}

\hypersetup{colorlinks,%
	citecolor=blue,%
	filecolor=green,%
	linkcolor=red,%
	urlcolor=violet,%
}

\newcommand{\hell}{{\mathscr{H}}}

\newcommand{\Real}{\mathbb{R}}
\newcommand{\DoA}{\mathcal{D}}

\newcommand{\bftheta}{{\boldsymbol{\theta}}}

\newcommand{\bfmu}{{\boldsymbol{\mu}}}
\newcommand{\bfgamma}{{\boldsymbol{\gamma}}}
\newcommand{\bfSigma}{{\boldsymbol{\Sigma}}}
\newcommand{\bfJ}{{\boldsymbol{J}}}

\newcommand{\bfD}{{\boldsymbol{D}}}
\newcommand{\bfX}{{\boldsymbol{X}}}
\newcommand{\bfa}{{\boldsymbol{a}}}
\newcommand{\bfb}{{\boldsymbol{b}}}
\newcommand{\bfI}{{\boldsymbol{I}}}
\newcommand{\bfy}{{\boldsymbol{y}}}

\newcommand{\bfzero}{{\boldsymbol{0}}}

\title{Empirical Bayes inference for the block maxima method}
\author{S. A. Padoan, S. Rizzelli}

\begin{document}

\maketitle
\begin{abstract}
The block maxima method is one of the most popular approaches for extreme value analysis with  independent and identically distributed observations in the domain of attraction of an extreme value distribution. The lack of a rigorous study on the Bayesian inference in this context has limited its use for statistical analysis of extremes. In this paper we propose an empirical Bayes procedure for inference on the block maxima law and its related quantities. We show that the posterior distributions of the tail index of the data distribution and of the return levels (representative of future extreme episodes) satisfy a number of important theoretical properties. These guarantee the reliability of posterior-based inference and extend to the posterior predictive distribution, the key tool in Bayesian probabilistic forecasting. Posterior computations are readily obtained via an efficient adaptive Metropolis-Hasting type of algorithm. Simulations show its excellent inferential performances already with modest sample sizes. The utility of our proposal is showcased analysing extreme winds generated by hurricanes in the Atlantic basin.
\end{abstract}

\section{Introduction}

Extreme Value Theory (EVT) provides a mathematical foundation to analyse extreme events regardless of the unknown underlying distribution. In this paper we focus on the univariate  {\it Block-Maxima} (BM) approach (e.g. \citealp{coles2001}, Ch. 1; \citealp{beirlant2006}, Ch. 5). Namely, we consider a dataset of $n$ values from an unknown distribution, which are then divided in $k$ blocks say of $m$ observations and from which $k$ maxima are computed. These, if suitably normalised, are asymptotically distributed according to the Generalised Extreme Value (GEV) distribution for an increasing block-size, provided that
some weak conditions are satisfied \citep[][Ch. 1]{dehaan+f06}.
EVT's supreme aim is to provide the basis for the statistical prediction of future extreme events, a vital task for risk management. In statistics, a robust approach to prediction is through the Bayesian paradigm, which naturally takes into account model uncertainty. 
Several Bayesian analyses of univariate BM have been proposed over time (see e.g., \citealp{coles2001}, Ch. 9; \citealp{beirlant2006}, Ch. 11 and the references therein).
Nevertheless, the potential of Bayesian inference does not seem to be fully exploited in earlier BM literature. 

The first motivation regards the asymptotic nature of the EVT. In real applications BM only approximately follow a GEV distribution, since the block-size is fixed.  Accordingly, on the one hand, end users are inclined to believe that a very large block-size is needed in order to trust a GEV based analysis, a requirement that may be incompatible with the sample size available in many applications. 
On the other hand, end users often analyse BM pretending that they are exactly distributed according to a GEV distribution, the so-called {\it vanilla} approach, without worrying too much that it is in fact a misspecified model for the data. 
A full picture on the reliability of approximate GEV-based inference has been missing for years. In fact, in the frequentist context, although basic estimators such as the Maximum Likelihood (ML) and Probability Weighted Moments (PWM) (\citealp{jenkinson1969statistics}, \citealp{hosking1985estimation}) have been introduced decades ago, for a long time the only available asymptotic theory was that derived under the unrealistic vanilla framework (\citealp{smith1985maximum}, \citealp{diebolt2008improving}, \citealp{b+s2017}, see also \citealp{dehaan+f06} and \citealp{beirlant2006}  and the references therein for other estimators). A proper theory for such estimators that takes the model misspecification issue into account is only recently experiencing a proper development, see \citet{ferreira2015block},  \citet{dombry15} and \citet{dombry19}. While a vanilla approach leads to the naive conclusion that those estimators are asymptotically unbiased, disregarding the block-size, recent theory reveals the potential for an asymptotic bias, whenever the block-size does not increase sufficiently fast enough along with the number of blocks, as is expected.  

The second motivation regards the complexity of the GEV class. 
Specifically, the GEV is a three parameter (location, scale and shape) superset of distributions, comprehending three different subfamilies, known as short-, light- and heavy-tailed, depending on the values of the shape parameter. The latter, known as the tail index, describes the tail heaviness of a distribution \citep[see][Ch.1]{dehaan+f06}. 
Notably, its sign affects the GEV support, thus the GEV is an {\it irregular} class.
Location and scale are fixed parameters under the unrealistic vanilla framework, while they are block-size dependent in the more realistic BM approach.
Accordingly, the elicitation of a prior distribution for such parameters, capable of taking the block-size into account, is not an easy task under a genuinely Bayesian approach. Moreover, the expression of the posterior distribution is not available in closed-form, 
therefore posterior-based inference relies on the generation of samples  from the posterior via Markov Chain Monte Carlo (MCMC) methods.
In this regard, the dependence among parameters entails that standard MCMC procedures can be computationally and statistically inefficient when the prior distribution is not appropriately specified.

In this article, we propose an Empirical Bayes (EB) procedure for statistical inference on the BM law and its related quantities. Our setting leverages on the GEV approximation but regarding it as a misspecified model for the data.
Then, our basic idea, and the most natural choice in our opinion, is to define an overall prior distribution consisting of a data-independent prior for the tail index and data-dependent prior distributions for the location and scale parameters, overcoming the difficulties of an orthodox prior selection.
Our EB method does not allow to obtain a genuine posterior distribution since it does not follow an authentic Bayesian formulation. In the sequel, we however use the term ``posterior'' instead of ``quasi-posterior'', for simplicity.
The first main contribution of this paper is to derive a solid theory for the proposed method to guarantee trustworthy inferences.
We give simple conditions on the prior distribution to obtain a consistent posterior distribution of the GEV parameters. 
In particular, we quantify the rate at which the posterior distribution concentrates around the true parameter values, since this is useful in understanding whether accurate inference is achievable in practice with the available sample size. 
We then establish that the posterior distribution is asymptotically close to a normal distribution. This result is particularly useful to: show that posterior credible intervals have good frequentist coverage probability (close to the nominal  confidence level); to derive simple approximate credible regions for the entire parameter vector, preventing practical complications in the derivation of such regions for multidimensional parameters.

Under the BM approach, prediction of future extreme values, potentially larger than those observed in the available dataset, can be pursued by computing the so-called  {\it return level} corresponding to the {\it return period} $T$, namely the value that is expected to occur or being exceeded in the future, every $T$ time periods. The second main contribution of this paper is to show that our EB method produces a posterior distribution of the return level which also satisfies the aforementioned theoretical properties: consistency with suitable contraction rates and asymptotic normality. Furthermore, as the prediction through the posterior predictive distribution is unquestionably a Bayesian paradigm strength, we also establish asymptotic results that guarantee the accuracy of posterior predictive inference.

The second main contribution of this paper is practical. Firstly, we provide simple concrete examples of prior distributions that meet the conditions required by our theory. Secondly, we propose to adopt a simple adaptive random-walk Metropolis-Hasting algorithm \citep{haario2001} for the calculation of the posterior distribution, since by embedding our proposed prior distributions it turns out to be computationally and statistically efficient. By an extensive simulation study we show that the empirical posterior distributions obtained via this sampling scheme comply with the theoretical results. Most importantly, we demonstrate that an accurate inference is achievable with moderate dimensions of block-size and blocks number, realistically available in the vast majority of applications.

Our methods and data have been incorporated into the {\tt R} package {\tt ExtremeRisks}, freely
available on CRAN. The online Supplementary Material contains technical details, further simulation results and all the proofs. The paper layout is the following. Section \ref{sec:background} explains in detail our statistical context and introduces our empirical Bayes method. Section \ref{sec:consistency} provides the asymptotic theory for the posterior distribution of the parameters of the GEV family, the return levels, and for the posterior predictive distribution. 
Section \ref{sec:algorithm} describes practical aspects of the posterior distribution computation. The finite sample performance of the proposed methods is examined via simulation in Section \ref{sec:simulation} and on annual maxima wind speed data  generated by Atlantic basin hurricanes in Section \ref{sec:realdata}.

%
\section{Background}\label{sec:background}
%

\subsection{Setting and notation}

Let $X_1,\ldots,X_m$ be independent and identically distributed (iid) random variables with common (unknown) distribution $F$ which is assumed to be in the domain of attraction of a distribution $G_\gamma$, where $\gamma\in\Real$ is the tail index, in symbols $F \in\DoA(G_\gamma)$. This means that there are sequences $a_m>0$ and $b_m \in \Real$ such that
\begin{equation}\label{eq:DoA_standard}
\lim_{m\to \infty} F^{m}(a_m x+b_m)=G_\gamma(x),
\end{equation}
for all continuity points of $G_\gamma$, where the latter is a GEV distribution. From a purely distributional view point  a GEV distribution function is of the form $G_{\bftheta}(x):=G_\gamma\left((x-\mu)/\sigma\right)$ for $x\in \mathcal{S}_{\theta}$, where  $\bftheta:=(\gamma,\mu, \sigma)^\top$ ranges over $\Theta:= \Real^2\times (0,\infty)$, while
$$
G_\gamma(z)= 
\begin{cases}
\exp\left(-
 \left( 1+\gamma
z \right)_+^{-1/\gamma} 
\right), \hspace{1em} \gamma \neq 0\\
\exp\left(
-\exp\left(
-z
\right)
\right), \hspace{3.1em} \gamma=0
\end{cases}
$$
and $\mathcal{S}_{\bftheta}
$ is the set $[ \mu-{\sigma}/{\gamma}, \infty) $ or  $\Real$ or  $(-\infty ,\mu-{\sigma}/{\gamma}]$,  if $\gamma>0$ or $\gamma=0$ or $\gamma<0 $, respectively, see \citet[][Ch. 1]{dehaan+f06} for details. Parameters $\mu$ and $\sigma$ are the location and scale, while $\gamma$ is informative on how heavy the distribution tail is. The corresponding density is $g_{\bftheta}(x)=g_\gamma((x-\mu)/\sigma)/\sigma$, with
\begin{equation}\label{eq:GEV_density}
g_\gamma(z)= 
\begin{cases}
(1+\gamma z)^{-1/\gamma-1}
\exp\left(-
\left( 1+\gamma z \right)^{-1/\gamma} \right),
 \hspace{1em} \xi \neq 0\\
\exp\left(
-\exp\left(-z \right)
	\right), \hspace{9.8em} \xi=0
\end{cases}.
\end{equation}
The vanilla approach simply assumes that data are exactly distributed according to $G_{\bftheta}$, while the BM method considered in this paper adopts a different data generating mechanism. Specifically, 
we consider  $X_1,\ldots,X_{n}$ iid random variables, $n=1,2,\ldots$, whose distribution satisfies $F\in\DoA(G_\gamma)$. We then assume that the $n$ variables can be split into $k\geq1 $ blocks of size $m\geq 1$, so that $n=mk$. 
The BM method is now concerned with the analysis of the $k$ block-maxima, where the $i$-th block maximum is given by
$$
M_{m,i}= \max(
X_{(i-1) m+1}, \ldots, X_{i m}),\quad i\in\{1,\ldots,k\},
$$
and whose distribution is $F^m$. By the domain of attraction condition in \eqref{eq:DoA_standard},  also known as {\it first-order} condition, the BM distribution can be {\it approximated} by the GEV distribution  $G_{\bftheta_{m}}$ with $\bftheta_{m}:=(\gamma,b_m,a_m)$, for a large enough given block-size $m$.
The first-order condition is equivalent to the condition
\begin{equation}\label{eq_1stOrderCond}
\lim_{m\to \infty} \frac{V(m x)-V(m)}{a(m)}=\frac{x^{\gamma}-1}{\gamma}=:Q_{G_{\gamma}}(1-e^{-1/x}), \quad \forall \, x>0,
\end{equation}
where $a$ is a positive function, $V(y):=F^{\leftarrow}(e^{-1/y})$, $y>0$, $Q_{G_{\gamma}}(p):=G_{\gamma}^{\leftarrow}(1-p)$, $p\in(0,1)$ and  $f^{\leftarrow}$ denotes the generalised left-continuous inverse of a non-decreasing right-continuous function  $f$ \citep[see][Ch. 1]{dehaan+f06}, on which basis a possible selection for the norming constants is $b_m=V(m)$ and $a_m=a(m)$.
Result \eqref{eq_1stOrderCond} is also useful because it allows to approximate an extreme quantile of the distribution $F$ via
\begin{equation}\label{eq:ExtremeQuantile}
Q_{F}(p):=F^{\leftarrow}(1-p)\approx Q_{G_{\bftheta_m}}(p_m):=b_m + a_m Q_{G_{\gamma}}(p_m),
\end{equation}
for a fixed small exceeding probability $p$ and for a large enough $m$, where $p_m=1-(1-p)^m$.
In this context, to study the asymptotic behaviour of some estimators (see e.g. the estimators in \citealp{dehaan+f06} Ch. 3,  \citealp{dombry15} and \citealp{dombry19}, to name a few), the following {\it second-order} condition has been introduced
\begin{equation}\label{eq_2ndOrderCond}
\lim_{t\to \infty}\frac{\frac{V(tx)-V(t)}{a(t)}-\frac{x^{\gamma}-1}{\gamma}}{A(t)}=H_{\gamma,\rho}(x), \quad \forall\, x>0,
\end{equation}
where $A(t)$ is a positive or negative function satisfying $A(t)\to0$ as $t\to\infty$ such that 
$|A|$ is a regularly varying function with index $\rho \leq 0$, named the second-order parameter, while $H_{\gamma,\rho}$ is a non-null function whose expression depends on $\gamma$ and $\rho$ \citep[see][Appendix B, for details]{de1996generalized, dehaan+f06}. These are useful in practice for the derivation of the non-negligible bias factor of an estimator, due to model misspecification. In particular, \citet{dombry19} use them for developing the asymptotic theory of ML estimation in the frequentist context. 

In this paper, we extend their results to the Bayesian paradigm.
We conclude this section with some notation used throughout the paper.
For any pair of probability measures $F,H$ over a Borel subspace of $\Real$, with Lebesgue densities $f,h$, we denote by
$$
\hell(f,h)^2=\int \left( \sqrt{f(x)}-\sqrt{h(x)}\right)^2 \diff x, 
$$
 the squared Hellinger distance. For a real valued function $f$ on $\mathbb{R}$, $f'$ and $f''$ denote its first and second derivative.
The operations $c\bfX$ and $\bfX/c$, where $\bfX$ is either a vector or a matrix and $c$ is a scalar are meant componentwise.  The class of Borel subsets of $\Real^d$, $d=1,2,\ldots$, and $\bfX_k$ that is asymptotically distributed according to $F$ are denoted by $\mathcal{B}(\Real^d)$ and $\bfX_k \,\dot{\sim}F$.
\subsection{Inference}\label{sec:inference}
We specify here the inferential setting used to establish the asymptotic theory for our EB method, presented in the next section. Assume that the sequence $X_1,\dots,X_n$ follow a specific distribution $F_0\in\DoA(G_{\gamma_0})$, with $\gamma_0>-1$. Let $M_{m,i}$,  $\, i=1, \ldots,k $, be the sequence of BM with joint distribution $\prod_{i=1}^k F_0^{m}(\cdot)$ and whose probability density function is denoted by $f_0^{(m)}$. We assume that both $m$ and $k$ go to infinity and, to avoid asymptotic results based on double limits, in the sequel we assume for simplicity that $m$ depends on $k$, say $m\equiv m_k$, and that $m\to\infty$ as $k\to\infty$.

For a large fixed $m$, we assume that the family $\{G^k_{\bftheta}$, $\bftheta=(\gamma,b_{m}, a_{m})^\top\in \Theta=(-1,\infty)\times \Real\times (0,\infty)\}$ is used as the misspecified statistical model for the sequence $M_{m,i}$,  $\, i=1, \ldots,k $.  Note that without loss of generality here and in the sequel we use the simplified notation $\bftheta$ in place of $\bftheta_{m}$ since the parameter space $\Theta$ does not depend on $m$ and, accordingly, the reference statistical model is the same no matter what the block-size is. Analogously, we also use the symbol $\bftheta_0$ in place of $\bftheta_{m,0}=(\gamma_0,b_{m,0},a_{m,0})^\top$.
We denote the likelihood function relative to the GEV misspecified class by
$L_k(\bftheta)= \prod_{i=1}^k g_{\bftheta}(M_{m,i})$, for all $\bftheta \in\Theta$. Given the GEV log-density,
$$
l_{\bftheta}(x)=\begin{cases}\log g_{\bftheta}(x),\quad
x \in \mathcal{S}_{\bftheta}\\
-\infty, \hspace{3em} \text{otherwise}
\end{cases},
$$ 
for all $\bftheta \in\Theta$, where $g_\bftheta$ is as in \eqref{eq:GEV_density}, then the log-likelihood is simply defined as
$l_k(\bftheta)=\log L_k(\bftheta)=\sum_{i=1}^k l_{\bftheta} \left(M_{m,i}\right)$.
Accordingly, we denote by
\begin{equation}\label{eq:score}
S_{k,\bftheta_0}^{\top}=\left(\frac{1}{\sqrt{k}}\sum_{i=1}^k \frac{\partial {l_\bftheta}}{\partial \theta_j } \left(M_{m,i} \right)\right)_{j=1,2,3}\Bigg|_{\bftheta=\bftheta_0},
\end{equation}
the score process vector of the  log-likelihood evaluated at $\bftheta_0$.

Similarly to \citet[][Section 4]{padoan2020consistency}, the empirical Bayes procedure we propose is based on an overall prior distribution on $\bftheta$ of the following general form
\begin{equation}\label{eq:data_dep_prior}
\pi_k(\bftheta)= \pi_{\text{sh}}(\gamma) \pi_{\text{loc}}\left( 
\frac{b_m-\widehat{b}_{m,k}}{\widehat{a}_{m,k}}
\right)\frac{1}{\widehat{a}_{m,k}} \pi_{\text{sc}}
\left( 
\frac{a_m}{\widehat{a}_{m}}
\right)\frac{1}{\widehat{a}_{m,k}}, \quad \forall\,\bftheta \in \Theta,
\end{equation}
where the prior distributions on the location and scale parameters are data-dependent. More precisely, $\pi_{\text{sh}}$, $\pi_{\text{loc}}$ and $\pi_{\text{sc}}$ are generic probability kernels for the shape $\gamma$, rescaled location $b_m$ and rescaled scale $a_m$ parameters, respectively, which are not depending on the data (see Section \ref{sec:consistency} for more details), while the rescaling terms
$\widehat{b}_{m,k}$ and $\widehat{a}_{m,k}$ are estimators of $a_{m}$ and $b_{m}$, respectively.  The corresponding EB posterior distribution of the parameters of GEV family is therefore defined by
$$
\Psi_k(B):= \frac{\int_B
{L_k}(\bftheta) \pi_k(\bftheta) \diff \bftheta
}{\int_\Theta{L_k}(\bftheta) \pi_k(\bftheta) \diff \bftheta}, \quad\forall B\, \in \mathcal{B}(\Theta),
$$
where $\mathcal{B}(\Theta)$ is the class of Borel-subsets of $\Theta$. In the next section we establish the asymptotic properties of such EB posterior distribution.

We point out that that our theory benefits from the asymptotic results on the  local reparametrised likelihood and score processes developed by \citet{dombry19}. To save space we refer to Section 3.1 of the supplement for a detailed account, while below we only outline the basic idea. 
We remind that $G_{\bar{\bftheta}_0}$ with $\bar{\bftheta}_0=(\gamma_0,0,1)^\top$ arises as the limit distribution for suitably normalised maxima, see  \eqref{eq:DoA_standard}. Accordingly, the likelihood theory in \citet{dombry19} is established focusing on a GEV likelihood function defined using normalised maxima as data and the reparametrization
\begin{equation}\label{eq:reparametrization}
\bar{\bftheta}=r(\bftheta):=(\gamma, \,  (b_m-b_{m,0})/a_{m,0}, \, a_m/a_{m,0})^\top,
\end{equation}
whose corresponding MLE sets then out to estimate $\bar{\bftheta}_0$. Nevertheless, the likelihood defined on normalised BM is linked to that of unnormalised BM and the asymptotic results can be rephrased for estimation of $\bftheta_0$ and related quantities. This is especially relevant when interest goes beyond inference on the tail behaviour, extending e.g. to extrapolation beyond observed levels, see Section \ref{subsec:prediction_return_levels}.
A similar reasoning applies to the posterior distribution, namely  $\bar{\Psi}_k(B)=\Psi_k(r(B))$, where $\bar{\Psi}_k$ is a posterior using normalised BM as data. Since in applications the goal is often both  to assess the tail heaviness and to predict future extreme values, in the next section we present the asymptotic results regarding $G_{\bftheta_0}$ and its related quantities, while postponing the technical discussion on $G_{\bar{\bftheta}_0}$ to the supplement.

\section{Theory for Empirical Bayes Inference}\label{sec:consistency}
\subsection{GEV distribution: posterior asymptotic properties}\label{subsec:consistency}

In order to establish our theory we exploit the important notion of {\it contiguity}, we refer to \citet[][Ch. 6]{vdv2000} for a general discussion and to \cite{choud04} for an application to approximate statistical modelling of time series. Let $\mathcal{F}_0^{k}$ and $\mathcal{G}_0^{k}$ be the joint probability measures of the normalised maxima $(M_{m,i}-\beta(m))/a_{m,0}$, $i=1,\ldots,k$, and of $k$ iid random variables with $G_{\gamma_{0}}$ distribution, respectively.
Here, the function $\beta: \mathbb{N}\to \Real$ generalises the recentering through $b_{m,0}$, see Section X of the supplement for details.
Let $E_k$ be any measurable set sequence. Then, $\mathcal{F}_0^{k}$ is said to be contiguous with respect to $\mathcal{G}_0^{k}$, in symbols $\mathcal{F}_0^{k} \, \triangleleft \, \mathcal{G}_0^{k}$, if $\mathcal{G}_0^{k}(E_k)=o(1)$ implies that $\mathcal{F}_0^{k}(E_k)=o(1)$, as $k\to \infty$. 

Establishing the asymptotic behaviour of the numerator in the posterior distribution formula is essential to obtain posterior contraction rates and asymptotic normality. This is done, in the standard setting where the statistical model is well specified, by showing the existence of tests for the true parameter, with type I and II error probabilities having a suitable decay rate \citep{vdv2000,ghosal2017}. Thanks to the contiguity result in Theorem \ref{theo:contiguity}, this method can be adapted to the nonstandard setting where the misspecified model $G_{\gamma_0}$ is considered in place of the unknown distribution $F^m_0$.
\begin{theo}\label{theo:contiguity}
Assume that the following conditions are satisfied: 
\begin{inparaenum}
\item \label{enu:conti} $V_0$ is twice differentiable, 
\begin{equation}\label{cond:secondvonMises}
A(t):=t V_0''(t)/V_0'(t)-\gamma_0+1
\end{equation}
converges to $0$ as $t\to \infty$ and $|A|$
is regularly varying of order $\rho\leq 0$;
\item \label{enu:bias} $\sqrt{k}A(m)\to \lambda\in\Real$ as $k\to \infty$;
\item \label{enu:bounded_densities} $(\beta(m)-V_0(m))/a(m)=O(A(m))$ and, if $\lambda\neq 0$, there is an integer $m'\geq1$ such that
$$
\max_{m \geq m'} \Vert a(m) f_0^{(m)}(a(m)\cdot + \beta(m))/g_{\gamma_0} \Vert_\infty < \infty.
$$
\end{inparaenum}
Then, 
$
\mathcal{F}_0^{k} \triangleleft \mathcal{G}_0^{k}.
$
\end{theo}
Condition \ref{enu:conti} requires that $F_0^{\leftarrow}(\exp(-1/t))$ is a smooth quantile function and the rate functions $A$ defined as in \eqref{cond:secondvonMises}, see \citet{deHaan96}, is essentially regularly varying. This condition is reasonably mild and is statisfied by standard models considered in the simulation study of Section \ref{sec:simulation}, see Section 1.1 of the supplement. Note that under assumption \ref{enu:conti}, the second order condition \eqref{eq_2ndOrderCond} is satisfied with rate functions $A$ as in \eqref{cond:secondvonMises} and $a(m)=mV_0'(m)$. In addition, as $k\to \infty$
\begin{equation*}\label{eq:hellrate}
\hell(f_0^{(m)}, g_{\bftheta_0})=O(A(m)),
\end{equation*}
which is a very important property to derive the asymptotic results for density estimation in Theorem \ref{theo:contraction} and for prediction in Section \ref{subsec:posterior_predictive}, see Section 3.4 of the supplement.
Condition \ref{enu:bias} is the standard assumption adopted in EVT to quantify the bias amount of estimators arising from misspecified extreme value models \citep[e.g.,][]{dehaan+f06}. Finally, condition \ref{enu:bounded_densities} essentially requires that for a large block-size $m$ the approximating GEV density does not vanish faster than the true unknown one near the endpoints. This condition, also used by \cite{padoan2020consistency} to study multivariate maxima, does not appear over restrictive, we have indeed verified that standard models considered in the simulation study of Section \ref{sec:simulation} satisfy it, see Section 1.2 of the supplement.

We next specify simple conditions on our proposed data-dependent prior $\pi_k$ with generic form in \eqref{eq:data_dep_prior}, on the basis of which we establish our asymptotic theory. 
\begin{cond}\label{cond:prior}
The densities  $\pi_{\text{sh}}$, $\pi_{\text{loc}}$ and $\pi_{\text{sc}}$ satisfy the following conditions:
\begin{inparaenum}
\item\label{cond:pish}  $\pi_{\text{sh}}$ is a positive and continuous on $(\gamma_0\pm \eta)$, for an $\eta>0$.
\item\label{cond:pisc} there is $\eta \in (0,1)$ and a integrable continuous function  $u_{\text{sc}}:\Real_+\to\Real_+$ such that
\begin{inparaenum}
%
\item[(b.1)]\label{posit} $\inf_{x \in [1\pm \eta]}\pi_{\text{sc}}(x)>0$;
\item[(b.2)]\label{piscbound} $\sup_{ t \in (1\pm \eta)}\pi_{\text{sc}}(x/t) \leq u_{\text{sc}}(x)$, for all $ x>0$.
\end{inparaenum}
\item\label{cond:piloc}
there is $\eta \in (0,1)$ and
an integrable continuous function 
$u_{\text{loc}}:\mathbb{R}\to\Real_+$ such that
\begin{inparaenum}
%
\item[(c.1)]\label{posit1} $\inf_{x \in [-\eta, +\eta]}\pi_{\text{loc}}(x)>0$;
\item[(c.2)]\label{piscbound1} $\sup_{ t_1 \in (1\pm\eta), t_2 \in (-\eta, +\eta)}\pi_{\text{loc}}((x-t_2)/t_1) \leq u_{\text{loc}}(x)$, for all $ x>0$.
\end{inparaenum}
\end{inparaenum}
Furthermore, the estimators $\widehat{a}_{m,k}$ of $a_{m,0}$ and $\widehat{b}_{m,k}$ of $b_{m,0}$ are such that:
\begin{inparaenum}
\item[(d)]\label{cond:estim}  $\widehat{a}_{m,k}/a_{m,0}=1+o_p(1)$ and $(\widehat{b}_{m,k}-b_{m,0})/a_{m,0}=o_p(1)$, as $k\to\infty$.
\end{inparaenum}
\end{cond}
Conditions \ref{cond:pish}--\ref{cond:piloc} are satisfied by most of the usual probability density kernels. Prior densities of the shape parameter with bounded support (e.g. uniform) are also allowed, as long as the latter contains $\gamma_0$ as an interior point. Condition (d) is satisfied by classical estimators, such as the ML \citep[][Theorem 2]{dombry15} and the PWM \citep[][Theorem 2.3]{ferreira2015block}. We are now ready to establish posterior contraction rates. 
\begin{theo}\label{theo:contraction}
Let $X_1,\ldots,X_n$ be iid random variables with distribution $F_0\in\DoA(G_{\gamma_0})$. Let $M_{m,i}$, $i=1,\ldots,k$ be the corresponding BM. 
Assume that Condition \ref{cond:prior} and the assumptions of Theorem \ref{theo:contiguity} are satisfied.  Let $C_k $ be a sequence of positive real numbers satisfying $C_k \to \infty$ and $C_k=o(\sqrt{k})$ as $k\to \infty$, and set $\epsilon_{k}=C_k/\sqrt{k}$, $k=1,2,\ldots$. 
Then, there exist constants $c_1>0$ and $c_2>0$, such that, with probability tending to $1$ as $k \to \infty$:
\begin{inparaenum}
\item \label{res:para_contr}  
$\Psi_k\left(\left\lbrace
\bftheta \in \Theta: \, \left\Vert\left( \gamma-\gamma_0, \frac{b_{m}-b_{m,0}}{a_{m,0}}, \frac{a_m}{a_{m,0}}-1 \right) \right\Vert_1 >\epsilon_k
\right\rbrace\right)\leq e^{-c_1 C_k^2}$; 
\item \label{res:hell_contr} 
$\Psi_k(\{\bftheta\in\Theta : \, \hell (g_\bftheta, f_0^{(m)})>{\epsilon_k}\})\leq e^{-c_2 C_k^2}$.
\end{inparaenum}
\end{theo}
A main implication of results \ref{res:para_contr} and \ref{res:hell_contr} in Theorem \ref{theo:contraction} is that the posterior distribution  of $\bftheta$, based on unnormalised BM, provides consistent estimation of the unknown parameter $\bftheta_0$ and unknown true BM density $f_0^{(m)}$, cumulating its mass in a neighbourhood of those. See Proposition 4.20 in Section 4.3 of the supplement for the explicit result on the posterior consistency. 
The result at point \ref{res:para_contr} gives a refined Bayesian analog of the MLE consistency result in \citet[][]{dombry15}.
It is due to the fact that with high probability  the posterior of $\bar{\bftheta}$, based on normalised BM, concentrates most of his mass on a ball centered at $\bar{\bftheta}_0$, whose radius $\epsilon_k$ decreases with $k$, while out of the latter the residual mass decreases (exponentially fast) as $k$ increases (see Section 5.2 of the supplement). 
From a practical view point, this allows to check whether accurate posterior-based inference on the original parameter $\bftheta_0$ is achievable with finite samples. 
In Section \ref{sec:simulation} we indeed asses the degree of posterior concentration via simulation 
showing that with several standard statistical models accurate inference can be obtained using moderate dimensions for $m$ and $k$. 
We recall that the true density $f_0^{(m)}$ of the unnormalised BM becomes (topologically) undistinguishable from $g_{\bftheta_0}$ as $m$ increases. Since the posterior distribution concentrates on a set of parameters $\bftheta$ such that $g_{\bftheta}$ is close to $g_{\bftheta_0}$ (in Hellinger metric), then this allows for the concentration result in point  \ref{res:hell_contr} (see Sections 4.1 and 5.2 of the supplement). 
 Its relevant theoretical implications for statistical prediction are highlighted in Section \ref{subsec:prediction_return_levels}.
Theorem  \ref{theo:contraction} is also important from a technical view point because is preparatory for establishing posterior asymptotic normality. 
In the sequel,  the $d$-variate normal distribution is denoted by $\mathcal{N}(\bfmu,\bfSigma)$, where $\boldsymbol{\mu}$ and $\boldsymbol{\Sigma}$ are the mean and covariance matrix, respectively. When $d=1$, we write $\mathcal{N}(\mu,\sigma^2)$. We denote by $\mathcal{N}(\,\cdot \,; \boldsymbol{\mu},\boldsymbol{\Sigma})$ the probability measure of such a distribution. Moreover, the Fisher information matrix of the distribution $G_{\bar{\bftheta}_0}$ is denoted by $\boldsymbol{I}_0$, see Section 3.1 of the supplement and \citet{prescott1980} for details. We recall that  $S_{k,\bftheta_0} $ is the score vector process given in \eqref{eq:score}.
\begin{theo}\label{theo:BvM}
Assume that the conditions of Theorem \ref{theo:contraction} are satisfied. 
Then, as $k \to \infty$
$$
\sup_{B \in \mathcal{B}(\Real^3)
} |{\Psi}_k( \{
\bftheta \in \Theta: \sqrt{k}(r(\bftheta)-\bar{\bftheta}_0) \in B
\}
)
-\mathcal{N}(B; \boldsymbol{I}_0^{-1}S_{k,\bftheta_0}, \boldsymbol{I}_{0}^{-1})
 |=o_p(1).
$$
\end{theo}	
This result establishes that the posterior distribution of $\bar{\bftheta}$, based on normalised BM, is asymptotically close to a normal distribution centred at $\bar{\bftheta}_0+k^{-1/2}\boldsymbol{I}_0^{-1}S_{k,\bftheta_0}$ and with covariance matrix $k^{-1}\boldsymbol{I}_{0}^{-1}$, as $k$ increases. In particular, $\boldsymbol{I}_0^{-1}S_{k,\bftheta_0}$ is also asymptotically normally distributed and behaves like the normalised MLE, i.e. $\sqrt{k}(r(\widehat{\bftheta}_k)-\bar{\bftheta}_0)\, \dot{\sim} \, \mathcal{N}(\lambda\, \boldsymbol{I}_0^{-1} \boldsymbol{b}, 
\boldsymbol{I}_0^{-1})$ as $k\to\infty$, where $\boldsymbol{b}$ is a bias vector term, see Section 3.1 of the supplement and \citet[][Theorems 2.1-2.2]{dombry19} for details. 
Accordingly, $k^{-1/2}\bfI_0^{-1}S_{k,\bftheta_{0}}$ is asymptotically negligible as $k\to \infty$ and the posterior is therefore asymptotically centered at $\bar{\bftheta}_0$, consistently with the previous findings.
Since the reparametrization $r(\bftheta)$ is obtained via the simple linear transformation in \eqref{eq:reparametrization}, the main practical implication of Theorem \ref{theo:BvM} is that also the posterior distribution $\Psi_k$ based on nonnormalised BM asymptotically resembles a normal distribution, say $\mathcal{N}(\bfmu_k,\bfSigma_k)$, with $\bfmu_k$ and $\bfSigma_k$ that are suitable linear transformations of the score vector and of the inverse Fisher information matrix, respectively. 
In turn, also the univariate marginal distribution $\Psi_{k,j}$ of the individual parameter $\theta^{(j)}$, i.e. the $j$-th component of $\bftheta$ for  $j=1,2,3$, obtained from $\Psi_k$, 
is asymptotically normal. Therefore, considering an asymmetric ($1-\alpha$)-credible interval defined by $Q_{\Psi_{k,j}}(1-\alpha/2)$ and $Q_{\Psi_{k,j}}(\alpha/2)$ for any $\alpha\in(0,1)$, where $Q_{\Psi_{k,j}}(p)$ is the ($1-p$)-quantile of the posterior distribution $\Psi_{k,j}$, with $j=1,2,3$, its frequentist coverage probability asymptotically agrees with the nominal level $1-\alpha$, whenever $\sqrt{k}A(m)=o(1)$, i.e. $\lambda =0$
and so $\boldsymbol{I}_0^{-1}S_{k,\bftheta_0}\dot{\sim} \mathcal{N}(\bfzero, \bfI_0^{-1})$, as shown in the next result. 
\begin{cor}\label{cor:mparametercoverage}
For any $\alpha\in(0,1)$, let $I_{k,\alpha}^{A}=[Q_{\Psi_{k,j}}(1-\alpha/2);\;Q_{\Psi_{k,j}}(\alpha/2)]$, for $j\in\{1,2,3\}$.
Under the assumptions of Theorem 	\ref{theo:quantileBvM}, if $\lambda=0$, as $k\to \infty$ 
$$
\mathbb{P}\left(\{\theta_0^{(j)} \in I_{k,\alpha}^A \}\right)= 1-\alpha+o(1),\quad j=1,2,3.
$$
\end{cor}

Theorem \ref{theo:BvM} can also be exploited to draw practical guidelines for constructing (approximate) credible Highest Posterior Density (HDP) regions.
The derivation of HPD regions is a complex task when the expression of the posterior distribution is not known in closed-form. This is especially true for multidimensional parameters, as pointed out e.g. on page  262 of \citet[][]{robert2007} and in the references therein.
Since $\Psi_{k,j}$, $j=1,2,3$, is asymptotically similar to a normal distribution, say $\mathcal{N}(\mu_{k,j},\sigma^2_{k,j} )$, then the latter can be used to define HPD intervals with ($1-\alpha$)-credible level. Note that, $\mu_{k,j}$ and $\sigma^2_{k,j}$ depend on the true unknown parameter $\theta_0^{(j)}$ and therefore they cannot be used for interval estimation in practice. However, we can replace them by the posterior mean $\widehat{\mu}_{k,j}$ and variance $\widehat{\sigma}^2_{k,j}$ and then use the symmetric interval $I_{k,\alpha}^{S}=[\widehat{\mu}_{k,j}-z_{\alpha/2}\widehat{\sigma}_{k,j};\;\widehat{\mu}_{k,j}+z_{\alpha/2}\widehat{\sigma}_{k,j}]$, where $z_{\alpha/2}$ is the ($1-\alpha/2$)-quantile of an univariate standard normal distribution. The posterior asymptotic normality result is particularly useful when the interest is in computing a HPD region for the true unknown parameter vector $\bftheta_0$.  Again, since $\Psi_k$ can be approximated by $\mathcal{N}(\bfmu_k,\bfSigma_k)$, then a HPD region with ($1-\alpha$)-credibility for $\bftheta_0$ is given by the random symmetric ellipsoid
\begin{equation}\label{eq:ellipsoid}
E_{k,\alpha}=\widehat{\bfmu}_k + \widehat{\bfSigma}_k^{1/2}\,\text{B}_2\left(\bfzero, \sqrt{\chi^2_{3,1-\alpha}}\right),
\end{equation}
where $\text{B}_p(\bfzero,r)$ is the closed $L_p$-norm-ball in $\Real^3$ whose center is the origin $\bfzero$ and the radius is $r$, $\chi^2_{3,1-\alpha}$ is the ($1-\alpha$)-quantile of the chi-squared distribution with $3$ degrees of freedom and $\widehat{\bfmu}_k$ and $\widehat{\bfSigma}_k$ are the posterior mean and covariance, since $\bfmu_k$ and  $\bfSigma_k$ are not available.
\subsection{Return levels: posterior asymptotic properties}\label{subsec:prediction_return_levels}

Prediction of extreme events through the BM approach can be achieved estimating the so-called return level corresponding to a prefixed return period $T$. For instance, suppose $X_1,\ldots,X_{m}$ describe the random behaviour of an observational phenomenon, and the sequence $M_{m,i}$, $i=1,\ldots,k$ with $m=365$ is hence representative of yearly maxima.
Then, the return level $x_T$ is the value that is expected to occur or being exceeded on average every $T$ years. 
From a probabilistic view point, it is the $(1-p)$-quantile of the distribution $F_0^{m}$, where $p=1/T$ is a small exceedance probability such that $p=1-F_{0}^{m}(x_T)$. The distribution $F_0$ is unknown in real applications, therefore, for any $p\in(0,1)$, statistical inference on the unknown  quantile $Q_{F^m_0}(p)$ can be based on the approximation given by the GEV quantile $Q_{G_{\bftheta_0}}(p)$, where in particular
\begin{equation}\label{eq:return_levels}
Q_{G_{\bftheta}}(p)=b_{m} + a_{m}\frac{(-\log(1-p))^{-\gamma}-1}{\gamma}.
\end{equation}
For any $p\in(0,1)$, the map $q: \Theta\to \Real: \bftheta \mapsto Q_{G_{\bftheta}}(p)$ is continuous. Thus, the posterior distribution $\Psi_k$ on the GEV parameter $\bftheta$ induces a posterior distribution on the GEV quantile $q$, which is given by  $\Omega_k:= \Psi_k \circ q^{-1}$. Next, we provide a series of asymptotic results on the posterior distribution $\Omega_k$, which guarantee the reliability of quantile-based inference. 
\begin{theo}\label{theo:quantile_contr}
Assume that the conditions of Theorem \ref{theo:contraction} are satisfied. Let $C_k $ be a sequence of positive real numbers satisfying $C_k \to \infty$ and $C_k=o(\sqrt{k})$ as $k\to \infty$, and set $\epsilon_{k}'=a_{m,0}C_k/(\sqrt{k}|b_{m,0}|)$, $k=1,2,\ldots$.  Then, for every $p<1-e^{-1}$, there is a constant $c>0$ such that, with probability tending to $1$  as $k \to \infty$: 
$$
\Omega_k \left(q \in \mathbb{R}: \, 
|q/Q_{F_0^m}(p) -1| > \epsilon_k'
\right) \leq e^{-c\,C_k^2}.
$$
\end{theo}
The result in Theorem \ref{theo:quantile_contr} implies that for any $p\in(0,1)$, the posterior distribution $\Omega_k$, based on unnormalised maxima, is consistent and allows therefore for increasingly accurate inference on the true unknown quantile $Q_{F^m_0}(p)$, as $k$ increases.
Unlike the result in Theorem \ref{theo:contraction}, in this one the contraction rate of $\Omega_k$ depends on the tail heaviness of $F_0$.
The lighter is the tail of $F_0$, the the narrower is the neighbourhood of $Q_{F_0^m}(p)$ on which $\Omega_k$ concentrates, since $a_{m,0}/b_{m,0}=\max(\gamma_0,0)+o(1)$ as $k \to \infty$.
Hence, in practice, with short- or light-tailed distributions accurate estimate of BM quantiles are achieved with smaller sample sizes than in the heavy-tailed case.
The next result shows that the quantile posterior distribution is also asymptotically normal.
\begin{theo}\label{theo:quantileBvM}
Assume that the conditions of Theorem \ref{theo:contraction} are satisfied. Assume that $p<1-e^{-1}$ and set $\varrho_k=|b_{m,0}|\sqrt{k}/a_{m,0}$. Then, there is a nonnull vector of constants $\boldsymbol{D}_0\in \mathbb{R}^3$ and a constant $v\in \Real$  such that, as $k \to \infty$
$$
\sup_{B \in \mathcal{B}(\Real)}
 \left|
\Omega_k\left(
q \in \Real : \varrho_k\left(\frac{q}{Q_{F_0^m}(p)}-1\right) \in B
\right)
-
\mathcal{N}(B; \bfD_0^\top \bfI_0^{-1}S_{k,\bftheta_0}+v, \bfD_0^\top \bfI_{0}^{-1}\bfD_0)
\right|
=o_p(1).
$$
\end{theo}
The exact expression of $\boldsymbol{D}_0$ and $v$ and all the details are provided in Section 5.6 of the supplement. The main practical implication of Theorem \ref{theo:quantileBvM} is that the posterior distribution $\Omega_k$ is asymptotically close to a normal distribution, say  $\mathcal{N}(\mu_k,\sigma^2_k)$.
Once more, a benefit of posterior asymptotic normality is that it allows to understand whether ($1-\alpha$)-credibility intervals, for $\alpha\in(0,1)$, defined using  $\Omega_k$ have coverage probability with asymptotic nominal level ($1-\alpha$), in the frequentist sense.
In this regard, when $\lambda=0$ then we have $v=0$ and $\boldsymbol{D}_0^\top \boldsymbol{I}_0^{-1}S_{k,\bftheta_0} \overset{d}{\to} \mathcal{N}(0,\boldsymbol{D}_0 \boldsymbol{I}_{0}^{-1}\boldsymbol{D}_0)$ as $k\to\infty$. 
Accordingly, the next result shows that the frequentist coverage probability of asymmetric credible intervals defined by $Q_{\Omega_k}(1-\alpha/2)$ and $Q_{\Omega_k}(\alpha/2)$, i.e. the $1-\alpha/2 $ and $\alpha/2$ quantiles of $\Omega_k$, asymptotically agrees with the nominal level  $1-\alpha$.
\begin{cor}\label{cor:quantilecoverage}
For any $\alpha\in(0,1)$, let $I_{k,\alpha}^{A}=[Q_{\Omega_k}(1-\alpha/2);\;Q_{\Omega_k}(\alpha/2)]$.
Under the assumptions of Theorem 	\ref{theo:quantileBvM}, if $\lambda=0$, as $k\to \infty$ 
$$
\mathbb{P}\left(\{Q_{F_{0}^m}(p) \in I_{k,\alpha}^A \}\right)= 1-\alpha+o(1).
$$
\end{cor}
Another benefit of the asymptotic normality of the posterior distribution $\Omega_k$ is that for large $k$ the approximation with $\mathcal{N}(\mu_k, \sigma_k^2)$ can be used to alternatively derive an approximate HPD set with ($1-\alpha$)-credibility for $Q_{F^m_0}(p)$, for a small $p$. In practice, likewise Section \ref{subsec:consistency} it is given by the symmetric interval $I_{k,\alpha}^{S}=[\widehat{\mu}_{k}-z_{\alpha/2}\widehat{\sigma}_{k};\;\widehat{\mu}_{k}+z_{\alpha/2}\widehat{\sigma}_{k}]$, where $\widehat{\mu}_{k}$ and $\widehat{\sigma}_{k}$ are the posterior mean and standard deviation of $\Omega_k$.

Finally we recall that, for a small $p\in(0,1)$, the extreme quantile $Q_{F_0}(p)$ can be approximated by the right-hand formula in \eqref{eq:ExtremeQuantile}, which is equal to that in \eqref{eq:return_levels} with $m$ in front to the logarithm. Moreover, the posterior distribution $\Psi_k$ induces a posterior distribution on the approximate extreme quantile $Q_{G_{\bftheta_0}}(p_m)$, 
which is given by $\tilde{\Omega}_k:= \Psi_k \circ \tilde{q}^{-1}$, where  $\tilde{q}: \Theta\to \Real: \bftheta \mapsto Q_{G_{\bftheta}}(p_m)$, with $p_m$ as in \eqref{eq:ExtremeQuantile}. Thus, as done above for return levels, we define an asymmetric ($1-\alpha$)-credible interval by $I_{k,\alpha}^{A}=[Q_{\tilde{\Omega}_k}(1-\alpha/2);\;Q_{\tilde{\Omega}_k}(\alpha/2)]$, where $Q_{\tilde{\Omega}_k}(p)$ is the ($1-p$)-quantile of the posterior distribution $\tilde{\Omega}_k$, and an approximate HPD set with ($1-\alpha$)-credibility by the symmetric interval $I_{k,\alpha}^{S}=[\widehat{\mu}_{\tilde{q};k}-z_{\alpha/2}\widehat{\sigma}_{\tilde{q};k};\;\widehat{\mu}_{\tilde{q};k}+z_{\alpha/2}\widehat{\sigma}_{\tilde{q};k}]$, where $\widehat{\mu}_{\tilde{q};k}$ and $\widehat{\sigma}_{\tilde{q};k}$ are the mean and standard deviation of the posterior distribution $\tilde{\Omega}_k$, for the inference of $Q_{F_0}(p)$.

\subsection{Posterior predictive distribution}\label{subsec:posterior_predictive}

One unquestionable strength of the Bayesian paradigm is to be able to carry out prediction through the use of the posterior predictive distribution, which incorporates uncertainty on the model. In a BM approach, 
given a sequence of iid maxima $M_{m,i}$, $i=1,\ldots,k$, the distribution of the unobservable random variable $M_{m,k+1}$ can be described by the posterior predictive distribution  
\begin{equation}\label{eq:predictive}
\widehat{G}_{k}(x)=\int_{\Theta} G_{\bftheta}(x)\diff \Psi_k(\bftheta), \quad \forall x\in\Real,
\end{equation}
where  the GEV model $G_{\bftheta}$ approximates the unknown BM distribution $F^m_0$. Its density
$$
\widehat{g}_{k}(x)=\int_{\Theta} g_{\bftheta}(x)\diff \Psi_k(\bftheta), \quad \forall x\in\Real,
$$
is the posterior predictive density.
For $p\in(0,1)$, let $Q_{\widehat{G}_{k}}(p):=\widehat{G}_{k}^{\leftarrow}(1-p)$ be the ($1-p$)-quantile relative to $\widehat{G}_{k}$. Then, the computation of $Q_{\widehat{G}_{k}}(1/T)$ provides an alternative approach to the estimation of the return level, relative to the return period $T$, discussed in Section \ref{subsec:prediction_return_levels}, and, in turn, to prediction of future extreme observations.  The next result establishes key properties for the posterior predictive distribution and density.
\begin{cor}\label{cor:pred_contract}
Assume that conditions of Theorem \ref{theo:contraction} are satisfied. Let $K\subset(0,1)$ be any compact set. Then,  as $k\to \infty$: 
\begin{inparaenum}
\item \label{res:preddistr} 
$\sup_{B \in \mathcal{B}(\mathbb{R})}|\widehat{G}_k(B)-F_0^m(B)| = O_p(\epsilon_k)$;
\item \label{res:predquant}
$\sup_{p \in K} 	
\left|\frac{Q_{\widehat{G}_{k}}(p)}{Q_{F_0^m}(p)} -1\right| =O_p\left( \epsilon_k
\frac{a_{m,0}}{|b_{m,0}|}
\right).
$
\item \label{res:preddens}	 $\hell(\widehat{g}_k, f_0^{(m)} )= O_p(\epsilon_{k})$;
\end{inparaenum}
\end{cor}
First of all, point \ref{res:preddistr} guarantees that the predictive distribution $\widehat{G}_k$ provides a good approximation (in total variation sense) to the distribution $F_0^m$ of a future block-maximum $M_{m,k+1}$, for large enough $m$ and $k$, and therefore provides a reliable basis for the computation of predictive intervals for $M_{m,k+1}$. See  \citet{kruger2021} for Bayesian approaches to probabilistic forecasting that are based on the posterior predictive distribution. The result in point \ref{res:predquant}, which establishes the asymptotic accordance between the predictive quantile function  $Q_{\widehat{G}_{k}}(p)$ and the true unknown quantile function  $Q_{F^m_0}(p)$ in the functional sense, is especially important for return levels estimation. Indeed, the predictive quantile function $Q_{\widehat{G}_{k}}(\cdot)$ is particularly appealing for return level estimation as it implicitly takes model uncertainly into account. Our result ensures that the future unknown return level trajectory $\{Q_{F^m_0}(1/T),\,T_a,\leq T\leq T_b\}$ for a certain future return period interval $[T_a, T_b]$ is accurately estimated by the predictive-based trajectory $\{Q_{\widehat{G}_{k}}(1/T),\,T_a,\leq T\leq T_b\}$. The interest in rare events can go beyond return levels, for example, probabilistic forecasts of continuous variables can take the form of predictive densities \citep{gneiting07}. Within the BM approach, extreme regions can be derived through the use of the BM density $f_0^{(m)}$, as for example the density sublevel sets \citep{quantile92}. Thanks to the result in point \ref{res:preddens}, $\hat{g}_k$ is an asymptotically accurate estimator of $f_0^{(m)}$, thus offering  a solid mathematical ground  for the construction of predictive regions for future BM.

\section{Computational aspects}\label{sec:algorithm}
We resort to a MCMC computational method for the empirical calculation of the posterior distribution $\Psi_k$, due to the lack of its explicit formula.
In particular, to draw samples from the posterior distribution $\Psi_k$ we use the Adaptive (Gaussian) random-walk Metropolis-Hastings (AMH) scheme discussed in \citet{garthwaite2016}, which is a special case of the AMH class of algorithms  introduced by \citet{haario2001}.

The main elements are the likelihood function defined in Section \ref{sec:inference} that is here denoted by $L_k(\bftheta|\bfy)$, where the data sample of maxima is denoted by $\bfy=(y_1,\ldots,y_k)$ to simplify the notation, and  the data-dependent prior $\pi_k(\bftheta)$ in \eqref{eq:data_dep_prior}.  A prior density $\pi_k$, satisfying Condition \eqref{cond:prior}, is not difficult to specify. A simple example is indeed easily given by taking in \eqref{eq:data_dep_prior} the product of the following densities:  $\pi_{\text{sh}}(\gamma)=(1-T_1(-1))^{-1}t_1(\gamma)\ind(-1<\gamma<\infty)$, where $t_\nu$ and $T_\nu$ are the Studen-$t$ density and distribution function, respectively, with $\nu$ degrees of freedom, $\pi_{\text{loc}}((b_m-\widehat{b}_{m,k})/\widehat{a}_{m,k})/\widehat{a}_{m,k}=\phi((b_m-\widehat{b}_{m,k})/\widehat{a}_{m,k})/\widehat{a}_{m,k}$, where $\phi$ is the univariate standard normal density and  $\widehat{a}_{m,k}$ and $\widehat{b}_{m,k}$ are the ML estimators of $a_{m,0}$ and $b_{m,0}$ and $\pi_{\text{sc}}(a_m/\widehat{a}_{m,k})/\widehat{a}_{m,k}=\xi(a_m; 1,\widehat{a}_{m,k})/\widehat{a}_{m,k}$, where $\xi(x;\alpha,\beta)$, $x>0$ is a Gamma density with rate $\alpha>0$ and scale $\beta>0$. Other simple choices can be readily obtained, but we hereafter focus on this basic option.

A short summary of the algorithm is as follows. The current state of the chain $\bftheta^{(j)}$ at time $j$ is potentially updated by the proposal $\bftheta'\sim h(\bftheta| \bftheta^{(j)})=\phi_3(\bftheta^{(j)}, \kappa^{(j)} \bfSigma^{(j)})$, where $\phi_d(\bfmu, \bfSigma)$ denotes a $d$-dimensional Gaussian density function with mean $\bfmu$ and covariance matrix $\bfSigma$. 
Following \citet{haario2001}, the proposal covariance matrix $\bfSigma^{(j)}$ is specified as
\begin{equation}
\label{eq:A_update}
\bfSigma^{(j+1)} =
\begin{cases}
(1 + [\kappa^{(j)}]^2 / j) \bfJ_3 , & j \leq 100 \\
\frac{1}{j-1} \sum_{s=1}^j (\bftheta^{(s)} - \tilde{\bftheta}^{(j)})(\bftheta^{(s)} - \tilde{\bftheta}^{(j)})^\top + ([\kappa^{(j)}]^2/j) \bfJ_3, & j > 100,
\end{cases}
\end{equation}
where $\bfJ_d$ is the $d$-dimensional identity matrix, $\tilde{\bftheta}^{(j)}=j^{-1}(\bftheta^{(1)}+\cdots+\bftheta^{(j)})$,
and $\kappa^{(j)} >0$ is a scaling parameter that affects the 
acceptance rate of proposed parameter values.
According to \citet{garthwaite2016} we adaptively update $\kappa$ using a Robbins-Monro process so that
\begin{equation}
	\label{eq:updatetau}
	\log \kappa^{(j+1)} = \log \kappa^{(j)} + a (\eta^{(j)} - \eta^*),
\end{equation}
where $a = (2\pi)^{1/2} \exp(\zeta_0^2/2) / (2\zeta_0)$ is  a steplength constant, $\zeta_0 = - 1/\Phi(\pi^*/2)$, and where $\Phi$ is the univariate standard Gaussian distribution function. The parameter $\eta^* $ is the desired overall sampler acceptance probability, which we specified as $\eta^*= 0.234$, according to \citet{gelman1997}.
\begin{algorithm}[b!]
\caption{Adaptive Random-Walk Metropolis-Hastings MCMC scheme}
\label{alg:algo_joint}
\textbf{Initialize:} Set $\bftheta^{(0)}$, $\kappa^{(0)}$ and $\Sigma^{(0)}$\;
\SetKwRepeat{REPEAT}{repeat}{until}
\For{$j = 0$ \To{} $M$}{
%
%
Draw proposal $\bftheta' \sim \phi_3(\bftheta^{(j)}, \kappa^{(j)} \Sigma^{(j)})$.\;
Compute acceptance probability 
$\eta = \min \left( \frac{L_k(\bftheta'|\bfy)\pi_k(\bftheta')}{L_k(\bftheta^{(j)}|\bfy) \pi_k(\bftheta^{(j)})}, 1\right)$\;
Draw $U_1 \sim \mathcal{U}(0,1)$. If $\eta > U_1$ then set $\bftheta^{(j+1)} = \bftheta'$ else set $\bftheta_1^{(j+1)} = \bftheta^{(j)}$\;
Update $\bfSigma^{(j)}$ according to \eqref{eq:A_update} \;
Update $\kappa^{(j)}$ according to \eqref{eq:updatetau} \;
}
\end{algorithm}
Given the symmetry of the proposal, i.e. $h(\bftheta'| \bftheta) = h(\bftheta | \bftheta')$, the acceptance probability of the update 
$\bftheta^{(j+1)}=\bftheta'$ reduces to
$$
\eta^{(j)} = \min \left( \frac{L_k(\bftheta'|\bfy)\pi_k(\bftheta')}{L_k(\bftheta^{(j)}|\bfy)\pi_k(\bftheta^{(j)})}, 1\right),
$$
otherwise we set $\bftheta^{(j+1)}=\bftheta^{(j)}$ with rejection probability $1-\eta^{(j)}$.
This algorithm is summarised in Algorithm~\ref{alg:algo_joint}. 
Finally, let $\bftheta_1^*,\ldots,\bftheta_N^*$ be a sample generated from the posterior $\Psi_k$, then, for any $p\in(0,1)$, a sample $q_1^*,\ldots,q_N^*$ from quantile posterior $\Omega_k$ is obtained by exploiting the relation between $\bftheta$ and $q$ given in \eqref{eq:return_levels}. Moreover an approximation of the posterior predictive distribution $\widehat{G}_k$ can be obtained via simulation by considering the Monte Carlo counterpart of formula \eqref{eq:predictive}, i.e. $\widehat{G}(x)\approx N^{-1}\sum_{i=1}^NG_{\bftheta_i^*}(x)$.

\section{Simulation study}\label{sec:simulation}
In this section, we assess for finite sample  sizes the behaviour of the posterior distributions $\Psi_k$, $\Omega_k$ and $\tilde{\Omega}_k$, computed via the MCMC method described in Section  \ref{sec:algorithm}, and the performance of the resulting inference. To this purpose we consider nine distributions: three in the domain of attraction of the Fr\'echet, Gumbel and reverse-Weibull families, respectively. In the first case we focus on the unit-Fr\'echet, standard-Pareto and Half-Cauchy. These distributions share the same tail index $\gamma=1$. In the second case, we focus on the standard Gumbel, Exponential with unit rate and Gamma with both shape and rate parameters equal to $2$. The tail index is $\gamma=0$ with these models.  Finally, in the third case we focus on the reverse-Weibull with shape parameter equal to $3$, the Beta with shape parameters equal to $1$ and $3$ and the Power-law class $F(x)=1-K(x^* - x)^\alpha$, where $x^*$,  $\alpha$ and $K$ are the end-point of the distribution, the shape parameter and a positive constant, respectively. In particular, we set $x^*=5$, $\alpha=3$ and $K=1/9$.  The tail index is $\gamma=-1/3$ with these models. In the sequel, we only report the results regarding the Half-Cauchy, Gamma and Power-law distributions (one for each domain of attraction) while those of the remaining models are deferred to the supplement, to save space.

Our theory in Section \ref{sec:consistency} holds as long as the condition \ref{enu:bias} in Theorem \ref{theo:contiguity} is satisfied.
The function $|A|$ is regularly varying of order $\rho\leq0$ and in particular it asymptotically behaves as $C m^\rho$ for most of the distributions
in the Fr\'echet domain of attraction, where $C\neq0$ and $\rho<0$. For example, with the unit-Fr\'echet, standard-Pareto and Half-Cauchy distributions $\rho=-\infty, -1, -2$  and 
condition  \ref{enu:bias} is then satisfied by setting $k=m^{-2\rho} (\log(m))^{-\alpha}$ for some $\alpha\in\Real$. However, with the other domains of attraction different formulas for $k$ may be needed. To avoid using different selections for $k$ depending on the distribution taken into account, for simplicity we put  $k=m/\sqrt{\log(m)}$ for all the distributions. With this setting, condition \ref{enu:bias} is satisfied by the considered distributions apart from the Gamma model.
In this way, we are able to verify whether the posterior distributions behaviour and the corresponding inference agree with theory of Section \ref{sec:consistency} but also if the results are still satisfactory when the theoretical condition is violated. We consider the following specific block-sizes $m=40, 60,109, 234$ which entail the numbers of blocks $k=20,30,50,100$, corresponding to the sample sizes $n=km=800,1800,5450,23400$.  
We consider $k$ and $m$ that are modest in size, with the first three cases representing similar values to those found in real applications. 
\begin{figure}[t!]
\centering
\includegraphics[width=1\textwidth]{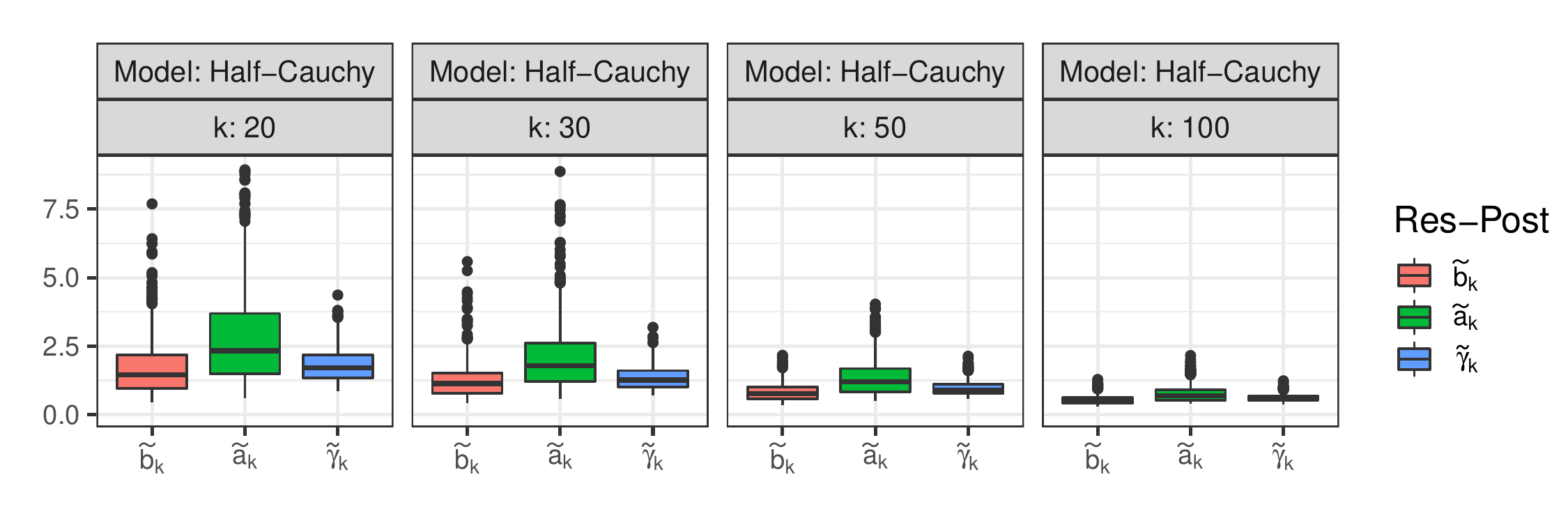}\\
\includegraphics[width=1\textwidth]{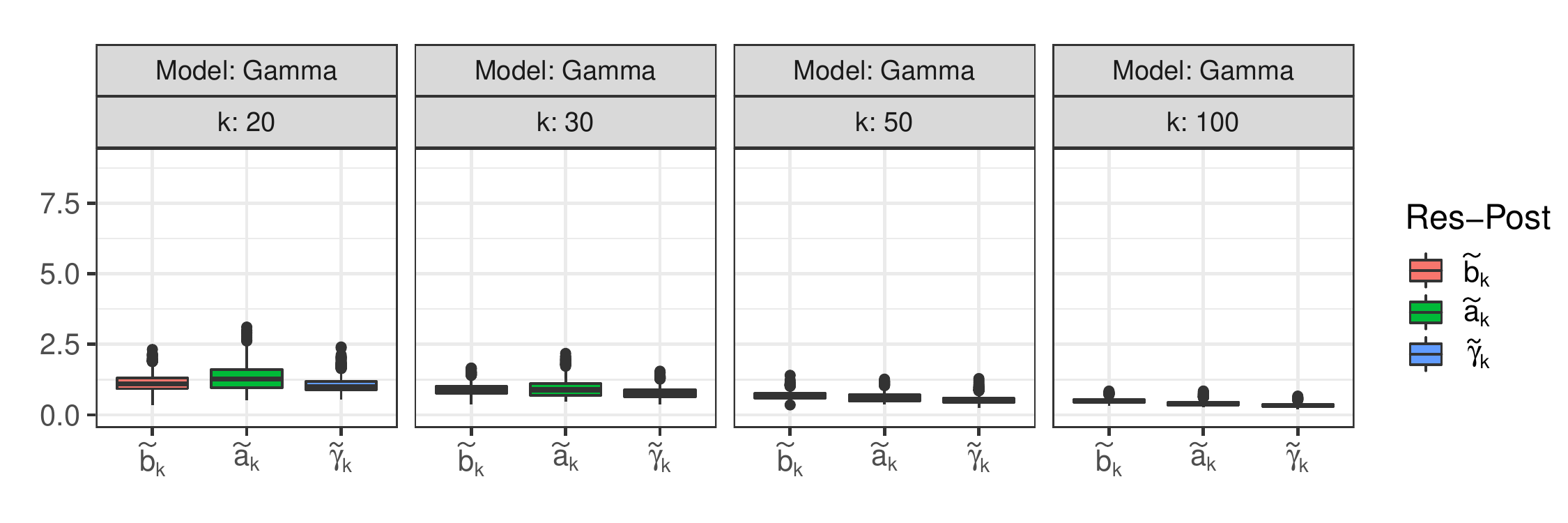}\\
\includegraphics[width=1\textwidth]{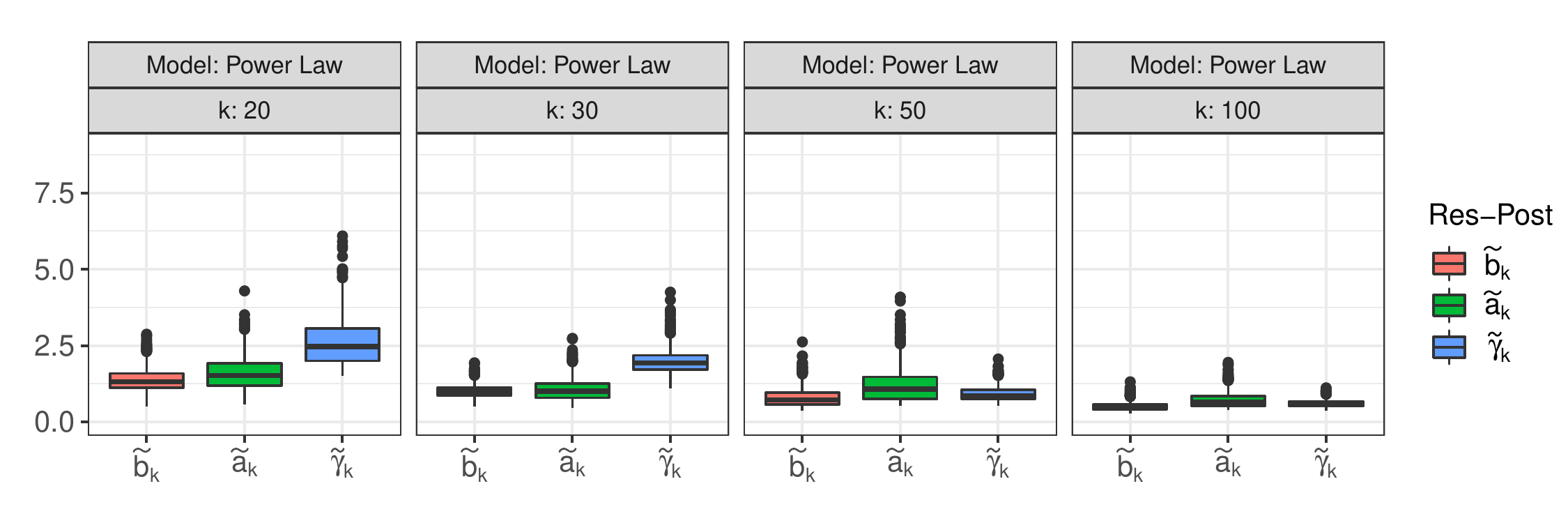}
\caption{Behaviour of the empirical posterior distribution of $\bftheta$ for increasing $k$. Box-plots of the Monte Carlo distribution of the summaries in \eqref{eq:posterior_statistics}.}
\label{fig:consistency}
\end{figure}
For each model we perform the following steps. In the first one, we simulate $n$ observations and derive on their basis $k$ maxima. Then, we run the AMH algorithm described in Section \ref{sec:algorithm} generating $50,000$ values of which we retain $N=20,000$ observations as a sample from the posterior $\Psi_k$, after a burn-in period of $30,000$. In the second step, we compute the following summaries. First, we calculate the uniform-norms 
\begin{equation}\label{eq:posterior_statistics}
\widetilde{\gamma}_k=\|\bfgamma^*_k-\gamma_0\|_\infty,\quad
\widetilde{b}_k=\|(\bfb^*_k-b_{m,0})/a_{m,0}\|_\infty,\quad
\widetilde{a}_k=\|(\bfa^*_k / a_{m,0} -1)\|_\infty,\quad
\end{equation}
where $\bfgamma^*_k=(\gamma^*_{k,1},\ldots,\gamma^*_{k,N})^\top$, $\bfb^*_k=(b^*_{k,1},\ldots,b^*_{k,N})^\top$ and $\bfa^*_k=(a^*_{k,1},\ldots,a^*_{k,N})^\top$ are the values sampled from $\Psi_k$, $(\gamma_0, b_{m,0}, a_{m,0})$ while are the true parameters with $a_{m,0}=mV_0'(m)$, $b_{m,0}=V_0(m)$. Then, we compute the proportion of times that the Manhattan-norm of the rescaled sampled values exceeds the radius $\epsilon_k$, i.e.
\begin{equation}\label{eq:_L1_exc_prop} 
\widetilde{p}_N=\frac{1}{N}\sum_{i=1}^N\ind\left(\left\|\left(\gamma^*_{k,i}-\gamma_0,
\frac{b^*_{k,i}-b_{m,0}}{a_{m,0}}, 
\frac{a^*_{k,i}}{a_{m,0}}-1\right)\right\|_1 > \epsilon_k\right),
\end{equation}
where $\epsilon_k=C_k/\sqrt{k}$ and $C_k=o(\sqrt{k})$ according to Theorem \ref{theo:contraction}. 
We set  $C_k=\log^2(k)$ obtaining $\epsilon_k\approx 2$  for all the $k$ values (see Table \ref{tab:_rate_conv}), which is a fairly small radius for all the considered models. In the third step, we compute: asymmetric and approximate symmetric  $95\%$-credibility intervals $I^{S}_{k,\alpha}$ and $I^{A}_{k,\alpha}$ for the individual true parameters $\gamma_0$, $a_{m,0}$, $b_{m,0}$, described in Section \ref{subsec:consistency}; the approximate symmetric $95\%$-credibility region for the true parameter vector $\bftheta_0$ defined in \eqref{eq:ellipsoid}; the asymmetric and approximate symmetric $95\%$-credibility intervals $I^{S}_{k,\alpha}$ and $I^{A}_{k,\alpha}$ for the return level $Q_{F^{m}_0}(1/T)$ with $T=100$, described in Section \ref{subsec:prediction_return_levels} and finally the asymmetric and approximate symmetric $95\%$-credibility intervals $I^{S}_{k,\alpha}$ and $I^{A}_{k,\alpha}$ for the extreme quantile $Q_{F_0}(p)$ of the underlying distribution $F_0$, with $p=0.001$, described at the end of Section \ref{subsec:consistency}. 

We repeat these three steps $M=1000$ times and with the obtained results we assess the following concentration properties of posterior distributions and coverage properties of posterior credible sets.  A sufficient condition for the posterior distribution to be consistent is that the theoretical counterpart of the summaries in \eqref{eq:posterior_statistics} converges to zero.
Figure \ref{fig:consistency} reports via box-plots the Monte Carlo  distribution of such summaries obtained with different models, along the rows, and different $k$ values, along the columns. In all the cases the right end point of the distribution is close to zero and the spread of the distribution is small already with $ k= 20$ (although some outliers are present). As $k$ increases, the range of the distributions shrinks considerably, becoming a very small interval in proximity to zero in the case $k=100$. This suggests that the posterior concentration property theoretically envisaged by consistency seems to hold in practice. Similar results are obtained also with other models, see Section 2 of the supplement.  
%
%
%
%
\begin{table}[t!]
\centering
{\small
\begin{tabular}{cccc|ccc|ccc|cccc}
\toprule
& & & & \multicolumn{3}{c}{Half-Cauchy} & \multicolumn{3}{c}{Gamma} &  \multicolumn{3}{c}{Power-law}\\
$k$ & $C_k$ & $\epsilon_k$ & $R(k)$ & $b_{m,0}$ & $a_{m,0}$ & $P_k$ & $b_{m,0}$ & $a_{m,0}$ & $P_k$ & $b_{m,0}$ & $a_{m,0}$ & $P_k$\\
\midrule
$20$ & $8.97$ & $2.01$ & $0.447$ & $25.8$ & $25.5$ & $94.0$ & $2.8$ & $0.58$ &  $99.9$ & $4.4$ & $0.20$ &  $100$\\
$30$ & $11.57$ & $2.11$ & $0.262$ & $38.5$ & $38.2$ &  $95.4$ & $3.0$ & $0.58$ &  $100$ & $4.5$ & $0.18$ &  $100$\\
$50$ & $15.30$ &$2.16$ & $0.096$ & $69.7$ & $69.4$ &  $96.6$ & $3.4$ & $0.57$ &  $100$ & $4.6$ & $0.14$ &  $100$\\
$100$ & $21.21$ &  $2.12$ & $0.011$ & $149.3$ & $149.0$ &  $98.3$ & $4.4$ & $0.56$ &  $100$ & $4.8$ & $0.08$ &  $100$\\
\bottomrule
\end{tabular}
}
\caption{Behaviour of the empirical posterior distribution of $\bftheta$ for increasing $k$. The seventh, tenth and thirteenth columns report the percentage in \eqref{eq:percentage} computed with the Student-$t$, Gamma and Power-law models and different values of $k$ and where $R(k)=e^{-0.01C_k^2}$.}
\label{tab:_rate_conv}
\end{table}
According to Theorem  \ref{theo:contraction}, the posterior distribution $\Psi_k$ places out of $\text{B}_1(\bfzero,\epsilon_k)$ (the $L_1$-norm-ball with centre $\bfzero$ and radius $\epsilon_k$) an amount of mass that goes to zero at the rate $R(k):=e^{-c\,C_k^2}$, for a certain $c>0$. 
Furthermore, the posterior mass outside of $\text{B}_1(\bfzero,\epsilon_k)$ is smaller than $R(k)$, with probability tending one.
%
%
%
\begin{table}[t!]
\centering
{\small
\begin{tabular}{lccc|cccccc}
\toprule
%
%
&  &  & & \multicolumn{6}{c}{Coverage probability}\\
\cdashlinelr{5-10}
Model & $k$ & $m$ & Type & $\gamma_0$ & $b_{m,0}$ & $a_{m,0}$ & $\bftheta_0$ & $Q_{F_{0}^m}(0.01)$ & $Q_{F_0}(0.001)$\\
\midrule
%
Half-Cauchy  &  $20$ & $40$ & S & $92.4$ & $96.1$ & $93.9$ & $92.1$ & $97.0$ & $95.0$\\
                     &           &          & A & $94.1$ & $94.5$ & $92.6$ &    --       & $93.6$ & $93.1$\\
\cdashlinelr{2-10}
                     & $30$ & $60$ &  S  & $94.9$ & $96.6$ & $95.4$ & $93.7$ & $97.1$ & $95.5$\\
                     &          &          &  A  & $94.8$ & $95.0$ & $94.4$ &       --   & $94.0$ & $94.1$\\
\cdashlinelr{2-10}
                     & $50$ & $109$ & S &$94.4$ & $95.9$ & $95.2$ & $93.8$ & $94.9$ & $94.6$\\
                     &          &            & A  &$93.7$ & $95.4$ & $94.2$ &    --       & $94.7$ & $94.7$\\
\cdashlinelr{2-10}
                    & $100$ & $234$ & S &$94.2$ & $95.4$ & $95.6$ & $94.9$ & $95.5$ & $96.2$\\ 
                    &            &            & A &$94.8$ & $95.0$ & $94.4$ &    --    & $94.7$ & $94.4$\\ 
\midrule
Gamma &  $20$ & $40$  & S & $96.5$ & $90.5$ & $93.4$ & $90.4$ & $97.1$ & $96.8$\\
              &           &           & A &$94.7$ & $90.8$ & $94.5$ &   --      & $94.6$ & $94.1$\\
\cdashlinelr{2-10}
               &  $30$ & $60$ & S & $96.4$ & $93.7$ & $93.4$ & $91.6$ & $96.8$ & $96.9$\\
               &           &          & A & $95.9$ & $93.6$ & $94.3$ &     --      & $94.8$ & $94.7$\\
\cdashlinelr{2-10}
               &  $50$ & $109$ & S & $95.9$ & $93.9$ & $93.0$ & $92.0$ & $96.5$ & $96.3$\\
               &           &            & A & $95.2$ & $94.2$ & $93.5$ &    --       & $95.2$ & $95.4$\\
\cdashlinelr{2-10}
               &  $100$ & $234$ & S & $95.5$ & $95.3$ & $94.3$ & $93.5$ & $96.8$ & $94.8$\\
               &             &            & A & $95.7$ & $95.5$ & $94.5$ &    --       & $95.3$ & $94.9$\\
\midrule
Power-law &  $20$ & $40$ & S & $96.5$ & $96.1$ & $97.7$ & $95.1$ & $96.7$ & $97.3$\\
                  &           &          & A & $96.3$ & $96.8$ & $97.3$ &    --      & $93.4$ & $93.5$\\
\cdashlinelr{2-10}
                  &  $30$ & $60$ & S & $95.5$ & $95.3$ & $96.7$ & $95.8$ & $97.0$ & $97.2$\\
                  &           &          & A & $94.2$ & $95.6$ & $96.3$ &    --       & $94.7$ & $94.8$\\
\cdashlinelr{2-10}
                  &  $50$ & $109$ & S & $95.4$ & $94.7$ & $95.5$ & $93.8$ & $95.3$ & $95.3$\\
                  &           &            & A & $94.4$ & $95.0$ & $94.3$ &     --      & $93.5$ & $94.2$\\
\cdashlinelr{2-10}
                  &  $100$ & $234$ & S & $95.4$ & $95.6$ & $96.0$ & $94.7$ & $96.6$ & $96.3$\\
                  &             &            & A & $95.0$ & $95.4$ & $95.5$ &     --      & $94.9$ & $96.1$\\
\bottomrule
\end{tabular}
}
\caption{Frequentist coverage probability of posterior credible region and intervals.}
\label{tab:_cov_prob_conf_int}
\end{table}
%
%
%
For practical purpose, in order to have an insight on the concentration speed of $\Psi_k$ around $\bftheta_0$, we have computed the percentage of the time that the proportion defined in \eqref{eq:_L1_exc_prop} is smaller than $R(k)$, where for the latter term we have set $c=0.01$, namely
\begin{equation}\label{eq:percentage}
P_{k}=\frac{1}{M}\sum_{i=1}^M\ind(\tilde{p}_{N,i}<R(k))\cdot 100\%,
\end{equation}
where $\tilde{p}_{N,i}$ is the $i$th Monte Carlo realisation of the proportion in \eqref{eq:_L1_exc_prop}.
Table \ref{tab:_rate_conv} collects the results. With the Half-Cauchy model and when $k=20$, the computed posterior distribution places more than approximately $45\%$ of its mass out of $\text{B}_1(\bfzero,2)$ only $6\%$ of the time. And so on decreasing, until the case $k=100$ where only $1.7\%$ of the times more than $1\%$ of the posterior mass lies out of $\text{B}_1(\bfzero,2)$. Even better results are obtained with the Gamma and Power-law models. Similar results are obtained with the other models, see Section 2 of the supplement. These results suggest that the computed posterior distribution starts to be fairly concentrated around $\bftheta_0$ already with $k = 20$ and then very rapidly increases its concentration degree as $k$ increases. Then, from the practical point of view, accurate posterior-based inference is achievable with inexpensive values of $m$ and $k$.

Finally we study the Monte Carlo frequentist coverage probability of the credible region and intervals discussed in Sections \ref{subsec:consistency} and \ref{subsec:prediction_return_levels}. The results for the different models and dimensions of $k$ and $m$ are reported along the raws of  Table \ref{tab:_cov_prob_conf_int}. 
When the column Type is equal to S the coverage probabilities of the approximate symmetric ($1-\alpha$)-credibility intervals $I_{k,0.05}^S$ for $\gamma_0$,  $a_{k,0}$, $b_{k,0}$, $Q_{F^m_0}(0.01)$, $Q_{F_0}(0.001)$ and of the approximate symmetric ($1-\alpha$)-credibility region $E_{k,0.05}^S$ for $\bftheta_0$, are reported. Otherwise, the coverage probabilities of the asymmetric ($1-\alpha$)-credibility intervals $I_{k,0.05}^A$ again for $\gamma_0$,  $a_{k,0}$, $b_{k,0}$, $Q_{F^m_0}(0.01)$ and $Q_{F_0}(0.001)$, are reported. 

%
Overall, all the coverage probabilities are close to the nominal level $95\%$ already with $k=20$ and then they get even closer as $k$ increases. This finding is consistent with the previous outcome, as expected.
In particular, the coverage of approximate symmetric and asymmetric credible intervals is almost the same when estimating the parameters $\gamma_0$, $a_{k,0}$, $b_{k,0}$. While, asymmetric credible intervals outperform the approximate symmetric ones when estimating $Q_{F_0^m}(0.01)$ and $Q_{F_0}(0.001)$. The smallest coverage levels (although still good) are obtained with the credible region $E_{k,0.05}^S$, but this is expected as estimating a three-dimensional parameter is harder than estimating a scalar one. Similar results are obtained with the other models, see Section 2 of the supplement. Overall, the take home message of this study is that in practice the good properties suggested by our asymptotic theory can be observed even when $m$ and $k$ are modest in size.

%
%
\section{Atlantic basin hurricanes wind speed analysis}\label{sec:realdata}
%
%
%
Impacts of wind speed and heavy rainfall generated by hurricanes are catastrophic on the affected areas. Such catastrophes have been experienced for example with Hurricane Katrina in August 2005 which caused 1,836 deaths and approximately \$125 billion in damage in the United States (US), where the Maximum Wind Speed (MWS) reached a peak of about 280 km/h. With a similar MWS in September 2017 Hurricane Maria caused 3,022 deaths and approximately \$90 billion in damage in US and Greater Antilles. Similar episodes have also been experienced with other Hurricanes: Delta in 2020 (MWS $\sim$225 km/h, 6 deaths; $\sim$ \$2 billion), Dorian in 2019 (MWS $\sim$295 km/h, 78 deaths; $\sim$ \$5 billion), Sandy in 2012 (MWS $\sim$185 km/h, 233 deaths; $\sim$ \$68.7 billion), etc., see National Hurricane Center (NHC)'s Tropical Cyclone Reports at {\tt https://www.nhc.noaa.gov/data/tcr/}.

Understanding how frequently certain strong wind speeds occur is importance for planning actions that can mitigate a hurricane's devastating consequences. To this aim, we analyse the MWS generated by hurricanes. In particular, we consider the Revised Atlantic Hurricane Database developed by the NHC that collects important information on hurricanes that have occurred in the Atlantic basin from 1851 to 2020, available on the website of the National Oceanic and Atmospheric Administration US federal agency, see {\tt https://www.noaa.gov/}. We then focus on the Maximum sustained surface Wind Speed (again, shortened to MWS) in km/h, that is, the 6-hour maximum of one minute average wind associated with the tropical cyclone at 10m elevation, and we derive and analyse the sequence of the largest MWS generated from hurricanes in a year and in that area, referred to as the annual MWS.
\begin{figure}
\centering
\includegraphics[width=0.24\textwidth, page=1]{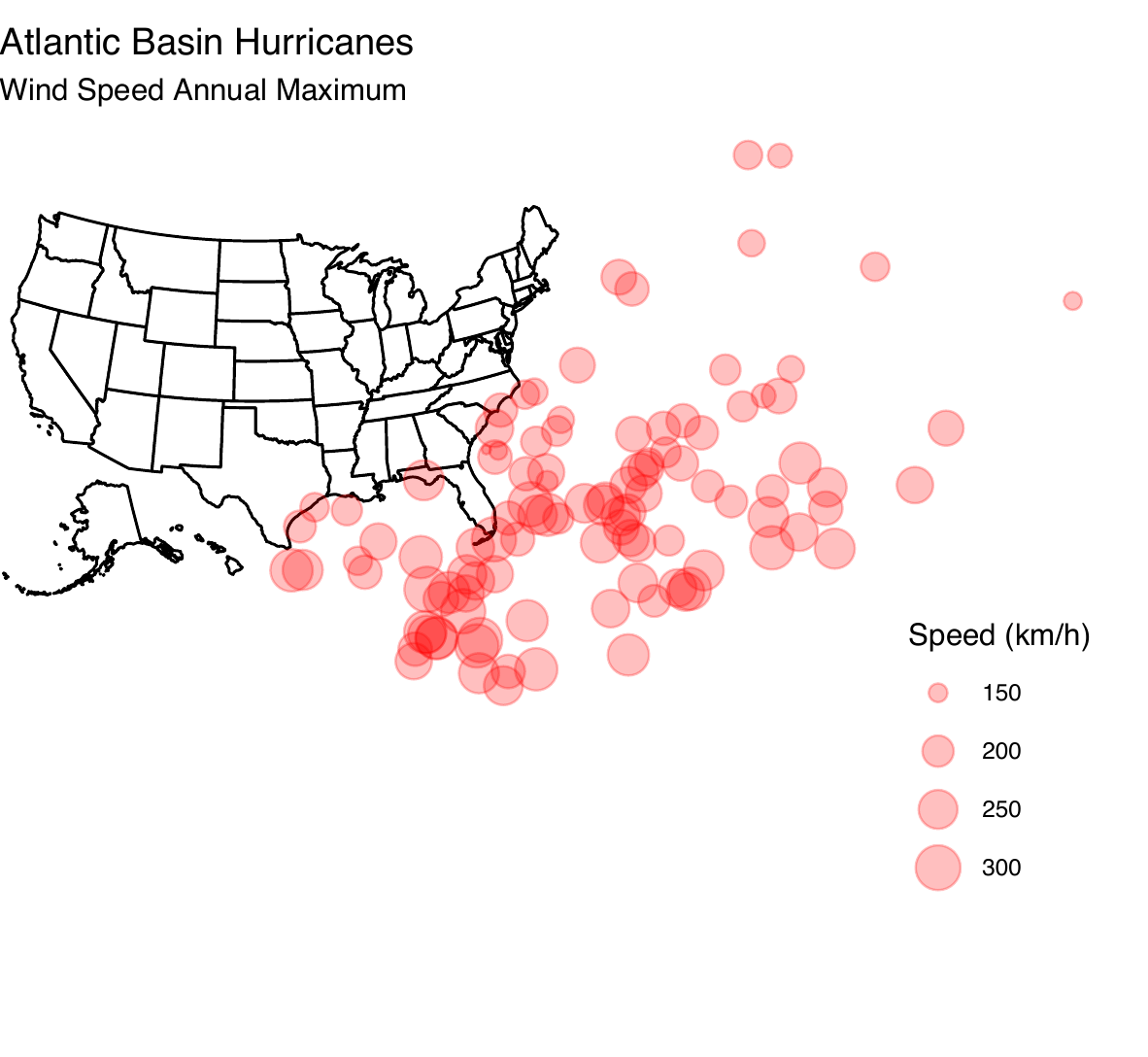}
\includegraphics[width=0.24\textwidth, page=2]{ABHurricane.pdf}
\includegraphics[width=0.24\textwidth, page=3]{ABHurricane.pdf}\\
\includegraphics[width=0.24\textwidth, page=4]{ABHurricane.pdf}
\includegraphics[width=0.24\textwidth, page=5]{ABHurricane.pdf}
\includegraphics[width=0.24\textwidth, page=6]{ABHurricane.pdf}\\
\includegraphics[width=0.24\textwidth, page=7]{ABHurricane.pdf}
\includegraphics[width=0.24\textwidth, page=8]{ABHurricane.pdf}
\includegraphics[width=0.24\textwidth, page=9]{ABHurricane.pdf}
\caption{Empirircal Bayes analysis of the annual MWS sequence recorded in the Atlantic basin from 1915 to 2020.}
\label{fig:return_levels_outcome}
\end{figure}
%
The database reports only the sequence of 6-hour maxima corresponding to the hurricane episodes, but as the strongest wind speed in a year occurs indeed in those circumstances, then to compute the annual MWS using the entire sequence of 6-hours maxima in a year is in fact the same as considering the shortened sequence. In a year, hurricanes take place at different times and in several far apart locations, spread in the Atlantic basin. Therefore, the overall sequence of 6-hour maxima consists of many independent observations and we then regard it as an iid sequence for simplicity.

We focus on the shortened annual MWS stationary sequence from 1915 to 2020, displayed in top-middle plot of Figure \ref{fig:return_levels_outcome}, as the entire series is not stationary due to a regime change between the series before 1915 and the one after. The series ranges between 140 km/h and 306 km/h and the  geographic locations at which the peaks were recorded are displayed in the top-left plot of Figure \ref{fig:return_levels_outcome}. The top-right panel reports the wind peaks histogram and estimated density, suggesting that the upper tail of the distribution is short.
%
\begin{table}[t!]
\centering
\caption{Maximum Likelihood Estimates (MLE), Posterior Mean (PM) and Asymmetric and Symmetric posterior 95\%-Credibility Intervals (95\%-A-CI and 95\%-S-CI) for parameters and return levels that are also forecasted by Posterior Predictive Quantiles (PPQ).}
{\small
\begin{tabular}{lccccccc}
\toprule
Parameter & MLE & PPQ& PM & 95\%-A-CI & 95\%-S-CI\\
\midrule
Location    & 216.7 & -- & 216.4 & [208.26,~224.12] & [208.19, 224.66]\\
Scale        & $~$37.3   & -- & $~$38.1   &  [ ~32.95, ~44.32] & [~ 32.21,~ 44.04]\\
Tail index  & $~~~$-0.35  & --&  $~~~$-0.35  & [ -0.458, -0.199]  & [ -0.470,  -0.203]\\
\midrule
2-years RL & 229.6 & 229.6 & 229.6 & [221.62, 237.45] & [221.55, 237.80]\\
5-years RL & 260.3 & 261.5 & 261.3 & [253.36, 269.55] & [253.38, 269.42]\\
10-years RL & 275.0& 276.8 &276.6 & [268.49, 285.89] & [268.22, 285.19]\\
15-years RL & 281.7& 283.8 & 283.7& [275.65, 294.22] & [274.63, 293.02]\\
\bottomrule
\end{tabular}
}
\label{tab:_rate_conv}
\end{table}
Nevertheless, the observed wind speeds are so extreme that forecasting future likewise severe or more severe events is still a crucial task for risks prevention. Indeed, according to the Saffir-Simpson Hurricane Wind Scale, a hurricane's maximum sustained wind speed is classified as category: 1 if between 119-153 km/h meaning that is very dangerous and will produce some damage; 2 if between 154-177 km/h meaning that is extremely dangerous and will cause extensive damage; 3 if between 178-208 km/h meaning that devastating damage will occur; 4 if between 209-251 km/h and 5 if higher than 252 km/h  both meaning that catastrophic damage will occur, see {\tt https://www.nhc.noaa.gov/aboutsshws.php} for details.
We computed the posterior distribution for the GEV parameters on the basis of the annual MWS sequence, applying the AMH algorithm described in Section \ref{sec:algorithm}. We generated 3,000 observations from the posterior (after 8k simulations and a 5k burn-in period). The posterior density of the location, scale parameters and tail index are reported from the middle-left to middle-right plots of Figure \ref{fig:return_levels_outcome}. Their shape is similar to that of a normal distribution, with that of the scale parameter that is a bit asymmetric on the right. 
The posterior mean, the asymmetric and symmetric  95\%-credibility intervals are reported in the fourth, fifth and sixth columns of the upper part of Table \ref{tab:_rate_conv}, respectively. In the second column the ML estimates are also reported. In particular, for the tail index, the posterior mean and posterior asymmetric (symmetric) 95\%-credibility interval, are approximately $-0.35$ and [-0.46, -0.20] ([-0.47, -0.20]), respectively, which support the hypothesis that the annual MWS follows indeed a short-tailed distribution, as expected. The posterior mean of the location parameter is 216 km/h which corresponds to a hurricane of category 4.
The bottom-left plot of Figure \ref{fig:return_levels_outcome} reports the posterior density of the 2-, 5-, 10- and 15-years return levels with the solid green, dashed orange, dotdash red and longdash magenta lines. The black dotted vertical line indicates the wind speed peak (280 km/h) reached by a category 5 hurricane such as Hurricane Katrina. The  fourth, fifth and sixth columns of the bottom part of Table \ref{tab:_rate_conv} report the mean, the asymmetric and symmetric  95\%-credibility intervals obtained from such posterior densities. These results suggest that a plausible return period for the wind speed 280 km/h (a category 5 hurricane) is 10 to 15 years, while in the shorter term such as 2 to 5 years such a speed is implausible (though not impossible in 5). These results are supported by the posterior predictive distribution ones whose forecasts on the  2-, 5-, 10- and 15-years return levels are indeed almost the same as the means of the posterior distribution, see the third column of Table \ref{tab:_rate_conv}. 
The bottom middle plot reports the estimated posterior predictive density (its mean - green square dot - and 5\% and 95\% quantiles - blue diamond dots).
The bottom right plot of Figure \ref{fig:return_levels_outcome} reports the forecasted return levels corresponding to a return period ranging from 2 to 1000 years. An estimate of the return levels is given by the posterior mean reported by the large black dots and the magenta solid (red dotted) lines report the bounds of the posterior asymmetric (symmetric) 95\%-credibility intervals. 
These suggest in particular that approximately 300 km/h
is a 50 years return level forecast, according to the posterior predictive distribution and also the posterior mean, and that an asymmetric (symmetric) posterior 95\%-credibility interval is approximately equal to [290 km/h, 315 km/h] ([287 km/h, 313 km/h]).
Then, a wind speed peak higher than that recorded with the Hurricane Katrina (horizontal orange dashed line) is approximately expected in 50 years. Moreover, the severity of wind speed peaks increases with the return period, thus recommending in conclusion that the necessary precautions must be taken in advance. Finally, the forecasts based on the posterior predictive distribution (blue dotdashed line) agree with those based on the posterior mean up to 110 years return period, after which the former are more pessimistic than the latter.

\section*{Acknowledgements}
The authors are grateful to Botond Szabo for his helpful suggestions. Simone Padoan is supported by the Bocconi Institute for Data Science and Analytics (BIDSA), Italy.

\bibliographystyle{chicago} 
\bibliography{bibliopm}

\end{document}